\documentclass[twocolumn,prl,superscriptaddress]{revtex4-1}
\usepackage{amsfonts}
\usepackage{amsmath}
\usepackage{amssymb}
\usepackage{color}
\usepackage{graphicx}
\usepackage{bm}
\usepackage{esint}
\usepackage{ulem}
\usepackage{xcolor}

\usepackage{soul}
\usepackage{cancel}
\usepackage{stmaryrd}

\begin{document}

\title{Critical fluorescence of a transmon at the Schmid transition}

\date{\today}
\author{M. Houzet}
\affiliation{Univ.~Grenoble Alpes, CEA, IRIG, Pheliqs, F-38000 Grenoble, France}
\author{L. I. Glazman}
\affiliation{Departments of Physics and Applied Physics, Yale University, New Haven, CT 06520, USA}
\begin{abstract}
We investigate inelastic microwave photon scattering by a transmon qubit embedded in a high-impedance circuit. The transmon undergoes a charge-localization (Schmid) transition upon the impedance reaching the critical value. Due to the unique transmon level structure, the fluorescence spectrum carries a signature of the transition point. At higher circuit impedance, {quasielastic} photon scattering may account for the main part of the inelastic scattering cross-section; we find its dependence on the qubit and circuit parameters.
\end{abstract}
\maketitle

{\it Introduction.--} A quantum-mechanical degree of freedom can be severely affected by its coupling to a dissipative environment. In a pioneering work~\cite{Schmid1983}, Schmid predicted that a superconducting circuit as elementary as a Josephson junction is insulating when it is ohmically shunted by a resistance larger than the resistance quantum, $R_Q=\pi \hbar/2e^2$. This result, which was further studied in Refs.~\cite{Bulgadaev1984,Guinea1985,Fisher1985}, reflects a charge localization transition, which is associated with the breaking of the ground state degeneracy. Remarkably the prediction holds at any ratio between the Josephson and charging energies of the junction, $\tilde E_J$ and $\tilde E_C=e^2/2\tilde C$, respectively, where $\tilde C$ is the junction capacitance~\cite{Schon1990}. So far evidence for the charge localization transition by dc~\cite{Yagi1997,Pentilla2001,Watanabe2003} or low-frequency~\cite{Murani2019} measurements remains elusive.

As a quantum many-body effect, the Schmid transition should not only affect the ground state, but the excited states as well. Here we find a spectroscopic signature of the transition in the fluorescence spectrum~\cite{footnote0} of a weakly nonlinear Josephson junction with $\tilde E_C\ll \tilde E_J$, a.k.a.~a transmon qubit~\cite{Koch2007,Schreier2008}, coupled to a Josephson-junction chain{. The chain} realizes a transmission line with an adjustable impedance~\cite{Masluk2012,Bell2012}, in which {plasmon} waves or, equivalently, microwave photons propagate freely. This setup attracted recent experimental interest~\cite{Puertas2019,Kuzmin2019,Leger2019} as a way to emulate quantum impurity problems with superconducting quantum circuits~\cite{Garcia-Ripoll2008,LeHur2012,Goldstein2013,Snyman2015,Gheeraert2018,Yamamoto2019}. Our theory predicts a characteristic dependence of the inelastic scattering cross-sections on the parameters of the setup. 

\begin{figure}
\includegraphics[width=0.9\columnwidth]{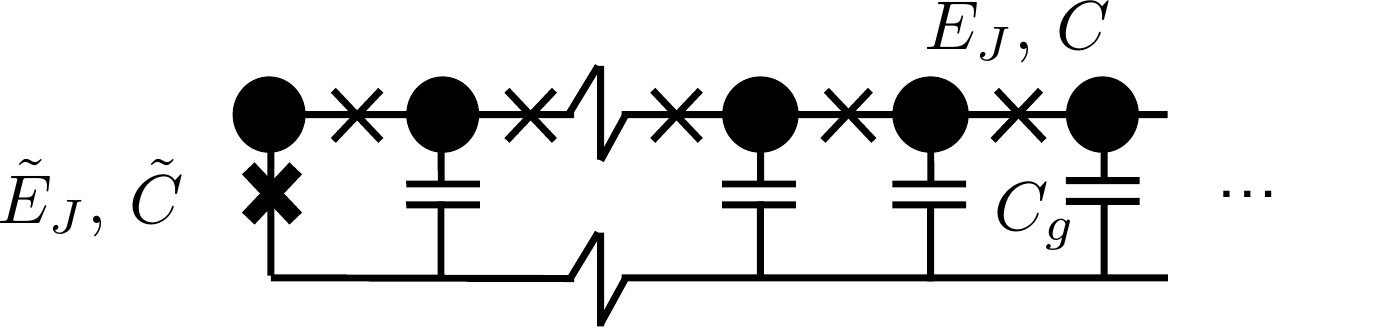}
\caption{\label{F:setup} 
{We study microwave photon scattering off a transmon coupled to a Josephson-junction chain. The transmon has Josephson energy $\tilde E_J$ and capacitance $\tilde C$. The junctions in the chain have Josephson energy $E_J$ and capacitance $C$, each superconducting island has ground capacitance  $C_g$.}
}
\end{figure}

{\it Model.--} Our consideration starts with the superconducting circuit of Fig.~\ref{F:setup}. The Josephson-junction chain to which the transmon is coupled is characterized by the Josephson energy $E_J$, and charging energies $E_C=e^2/2C$ and $E_g=e^2/2C_g$, where $C$ and $C_g$ are the chain's junction and ground capacitances. Under the conditions
\begin{equation}
\label{eq:condition}
E_CE_J\gg {\tilde E}_C{\tilde E}_J\gg {\tilde E}^2_C,
\end{equation}
the classical analysis, in which the Josephson junctions are approximated as linear inductances, yields a narrow transmon resonance that lies deep in the linear part of the waves' dispersion. The transmon resonance frequency and half-width are, respectively, $\omega_0=\sqrt{8{\tilde E}_C{\tilde E}_J}/\hbar$ and $\Gamma=1/2Z\tilde C$; the waves' dispersion is $\omega_p=vp/\sqrt{(vp/\omega_{\mathrm{B}})^2+1}$. Here $Z=R_Q/2K$ with $K=\pi\sqrt{E_J/8E_g}$ is the chain's impedance at low frequency, $\omega_{\mathrm{B}}=\sqrt{8{E}_C{E}_J}/\hbar$ is the photon bandwidth, $v=a\sqrt{8E_JE_g}/\hbar$ is the photon velocity and $a$ is the chain's unit cell length. Noting that $\Gamma=(4/\pi\hbar)K{\tilde E}_C$ and assuming $K\lesssim 1$ in the relevant range of parameters, we find from Eq.~\eqref{eq:condition} that, indeed, $\Gamma\ll \omega_0\ll \omega_{\mathrm{B}}$. Within the classical analysis, photon scattering is purely elastic.

To analyze inelastic scattering, we first notice that {$K=1/2$ is the critical value for the Schmid transition}~\cite{Schmid1983}. Below that value, the environment induces charge localization. {The elementary processes responsible for it are phase slips~\cite{Hekking1997}.} In an isolated transmon, the phase slip amplitude for the first excited band is~\cite{Koch2007}
\begin{equation}
\lambda_1=\frac{64}{\sqrt{\pi}}\frac{{\tilde E}_C}{\hbar} \left(\frac{2\tilde E_J}{\tilde E_C}\right)^{5/4}e^{-\sqrt{ 8{\tilde E_J}/{\tilde E_C}}}
\label{eq:lambda1}
\end{equation}
(parametrically larger than the phase slip amplitude for the ground state, $\lambda_1/\lambda_0=-8\sqrt{2{\tilde E}_J/{\tilde E}_C}$). Therefore, $\Gamma$ may become parametrically smaller than $\lambda_1$ {\it only} deep in the localized regime, $K\ll 1$. {Barring that, we} {treat phase slips perturbatively}. In most of the discussion below we assume $\lambda_1\ll\Gamma$. At the same time, the proliferation of phase slips in the chain's junctions, with an amplitude $\lambda_{\rm chain}\propto e^{-\sqrt{8E_J/E_C}}$, is known to drive a superfluid-to-insulating transition below the critical value $K=2$~\cite{Bradley1984,Korshunov1989,Basko2020}. That physics can be disregarded either if {the bandwidth $\Gamma$ that limits frequency exchange in a quasielastic process} (see below) is larger than the insulating gap, $\Gamma\gg\lambda_{\mathrm{chain}}$, or if the chain is short enough to evade the thermodynamic limit, $L<v/\lambda_{\rm chain}$ (but still long enough to ignore effects related to a finite level spacing, $\Delta=\pi v/L\to 0$). 

Ignoring the chain's nonlinearity we use Hamiltonian
\begin{equation}
\label{eq:H0}
H=4\tilde E_C(\hat{N}-\hat{n})^2+\tilde E_J[1-\cos{\hat\varphi}]+\sum_p\hbar \omega_pa^\dagger_pa_p
\end{equation}
to describe the setup. Here $\hat\varphi$ and $\hat N$ are the transmon's conjugate phase and charge operators, $a_p$ and $a^\dagger_p$ are the annihilation and creation operators of a {linearly-dispersing} photon of wave vector $p$, and
\begin{equation}
\label{eq:charge-displacement}
\hat{n}=\frac 1 \pi \sum_{p>0 }f_p(a_p+a_p^\dagger) \quad \mathrm{with}\quad f_p=\sqrt{ {K\Delta}/{\omega_p}}
\end{equation}
is the charge displacement operator at the qubit. The sum in Eq.~\eqref{eq:charge-displacement} includes only the dynamical variables of the chain. The static mode ($p = 0$) compensates for an eventual charge offset, which is effectively attenuated by the total capacitance of the array, $\sim C_g\!\cdot\! (L/a)$~\cite{vlad}.

{
The low-energy properties of Eq.~\eqref{eq:H0} are described by the boundary sine-Gordon model, $H_{\rm sG}=\sum_p\hbar \omega_pa^\dagger_pa_p-\lambda_0\cos(2\pi \hat{n})$~\cite{Gogolin-book}. Keeping the ratio $\tilde E_C/\tilde E_J$ small but finite generalizes it to a new quantum impurity problem, in which the transmon resonance brings in a nontrivial structure of the high-frequency spectrum, as we discuss now.}


{\it Phase-slip vs quartic anharmonicities.--} We first show, by perturbative {analysis in coupling $f_p$,} that phase slips play the dominant role in {quasielastic (or soft) photon scattering, i.e., inelastic scattering with a small energy transfer between the incoming and one of the outgoing photons.} For this we write the Hamiltonian \eqref{eq:H0} in the transmon basis of discrete eigenstates (we set $\hbar=1$ hereinafter),
\begin{eqnarray}
H&=&H_0+V,\quad H_0=\sum_s\varepsilon_s|s\rangle\langle s|+\sum_p\omega_pa^\dagger_pa_p+4E_C{\hat{n}}^2,\nonumber\\
V&=&-\hat{n}\sum_{ss'}W_{ss'}|s\rangle\langle s'| \quad {\rm with} \quad W_{ss'}=8E_C\langle s|\hat{N}|s'\rangle.
\label{eq:H}
\end{eqnarray}
Here we ordered the transmon eigenenergies, $\varepsilon_s>\varepsilon_{s'}$ if $s>s'\geq 0$, and set $\varepsilon_0=0$. The partial inelastic cross-section for a photon with frequency $\omega$ to be converted into three photons with frequencies $\omega_1,\omega_2,\omega_3$, such that $\omega=\omega_1+\omega_2+\omega_3$, 
is obtained {with Fermi's Golden Rule}, 
\begin{equation}
\label{eq:partialCross}
\gamma(\omega_1,\omega_2,\omega_3|\omega)=\frac{2\pi^2}{3!}\frac{\left|{\cal A}_{p;p_1p_2p_3}\right|^2}{\Delta^4}
\end{equation}
($3!$ accounts for permutations of momenta that describe the same final state) with $\omega=\omega_p$, $\omega_i=\omega_{p_i}$, and {a matrix element obtained perturbatively in ${f_p}$,}
\begin{eqnarray}
\label{eq:amplitude}
{\cal A}_{p;p_1p_2p_3}&=&\langle 0| a_{p_1}a_{p_2}a_{p_3}V\left(\frac 1{\omega_{{p}}-H_0}V\right)^3a^\dagger _p|0\rangle\nonumber\\
&=&-\frac{K^2\Delta^2}{\pi^4\sqrt{\omega\omega_1\omega_2\omega_3}}\sum_{srt}W_{0s}W_{sr}W_{rt}W_{t0}
\nonumber\\
&&\times\left\{
\frac 1{(\varepsilon_s+\omega_1+\omega_2-\omega)(\varepsilon_r+\omega_1-\omega)(\varepsilon_t-\omega)}
\right.
\nonumber\\
&&+\left.
\frac 1{(\varepsilon_s+\omega_1+\omega_2-\omega)(\varepsilon_r+\omega_1-\omega)(\varepsilon_t+\omega_1)}
\right.
\nonumber\\
&&+\left.
\frac 1{(\varepsilon_s+\omega_1+\omega_2-\omega)(\varepsilon_r+\omega_1+\omega_2)(\varepsilon_t+\omega_1)}
\right.
\nonumber\\
&&+\left.
\frac 1{(\varepsilon_s+\omega_1+\omega_2+\omega_3)(\varepsilon_r+\omega_1+\omega_2)(\varepsilon_t+\omega_1)}
\right.
\nonumber\\
&&+\left.\text{permutations of}\,\, \omega_1,\omega_2,\omega_3\,\right\}.
\end{eqnarray}
Here we ignored the last term in $H_0$ (it vanishes in the thermodynamic limit), {and the {summation is over} the transmon levels}. The most divergent terms correspond to the following sequences of transmon virtual states,
$0\shortrightarrow 1\shortrightarrow 0\shortrightarrow t\shortrightarrow 0$, 
$0\shortrightarrow 1\shortrightarrow r\shortrightarrow 1\shortrightarrow 0$, or
$0\shortrightarrow s\shortrightarrow 0\shortrightarrow 1\shortrightarrow 0$,
and to frequencies such that
\begin{equation}
\label{eq:condition2}
|\omega-\varepsilon_1|,|\omega_1-\varepsilon_1|,\omega_2,\omega_3\ll\varepsilon_1
\end{equation}
(as well as two other inequalities after permutation of $\omega_1,\omega_2,\omega_3$). The contribution to ${\cal A}_{p;p_1p_2p_3}$ that corresponds to the on-shell condition, $\omega=\omega_1+\omega_2+\omega_3$, and {satisfies} the inequalities \eqref{eq:condition2}, is
\begin{equation}
\label{eq:contribution}
-\frac{2K^2\Delta^2}{\pi^4\varepsilon_1\sqrt{\omega_2\omega_3}}\frac{|W_{01}|^2}{(\omega-\varepsilon_1)(\omega_1-\varepsilon_1)}\sum_s\left(\frac{|W_{1s}|^2}{\varepsilon_s-\varepsilon_1}-\frac{|W_{0s}|^2}{\varepsilon_s}\right).
\end{equation}
By substituting the operator $\hat{n}$ with a gate charge $\cal N$ in Eq.~\eqref{eq:H}, one can calculate the gate sensitivity of the energy levels $\varepsilon_{1,0}({\cal N})$ perturbatively in $\cal N$, and identify~{\cite{footnote7}}
\begin{equation}
\label{eq:phase-slip}
\sum_s\left(\frac{|W_{1s}|^2}{\varepsilon_s-\varepsilon_1}-\frac{|W_{0s}|^2}{\varepsilon_s}\right)
=\frac 12\left.\frac{\partial^2[\varepsilon_1({\cal N})-\varepsilon_0({\cal N})]}{\partial{\cal N}^2}
\right|_{{\cal N}=0}.
\end{equation}
In the transmon limit [cf.~Eqs.~(\ref{eq:condition}) and (\ref{eq:lambda1})], $\varepsilon_s({\cal N})=\varepsilon_s(1/4)+\lambda_s\cos(2\pi {\cal N})$, allowing us to replace the right-hand side of the exact 
Eq.~\eqref{eq:phase-slip} with $2\pi^2(\lambda_1-\lambda_0)\approx 2\pi^2\lambda_1$. Evaluating all remaining factors in Eq.~\eqref{eq:contribution} within the transmon's harmonic approximation, in which $\varepsilon^{(0)}_s\approx s\omega_0$ and $W_{sr}^{(0)}=8\tilde E_C(\tilde E_J/32\tilde E_C)^{1/4}[\sqrt{s}\delta_{s,r+1}+\sqrt{s+1}\delta_{s,r-1}]$, we find the leading contribution to Eq.~\eqref{eq:partialCross} for frequencies satisfying the conditions \eqref{eq:condition2},
\begin{equation}\label{eq:gamma3}
\gamma_{\rm si}
(\omega_1,\omega_2,\omega_3|\omega)=\frac{16}{3}\frac{K^2\Gamma^2\lambda_1^2}{\omega_2\omega_3(\omega_1-\omega_0)^2(\omega-\omega_0)^2}.
\end{equation}
The partial cross-section of {quasielastic} scattering, {which characterizes the transmon fluorescence,}
 \begin{equation}
 \label{eq:partialCross2}
\gamma_{\rm si}
(\omega'|\omega)=\int_0^{\omega-\omega'}d\omega_2\gamma_{\rm si}
(\omega',\omega_2,\omega-\omega'-\omega_2|\omega)\,,
\end{equation}
is obtained at $|\omega-\omega_0|,|\omega'-\omega_0|\ll \omega_0$ from Eq.~\eqref{eq:gamma3} and two other equations with permutations of $\omega_1,\omega_2,\omega_3$, as
 \begin{equation}
 \label{eq:gamma3-bis}
\gamma_{\rm si}(\omega'|\omega)=\frac{32K^2\Gamma^2\lambda_1^2}{(\omega-\omega_0)^2(\omega'-\omega_0)^2(\omega-\omega')}\ln\left(\frac{\omega-\omega'}\Delta\right);
\end{equation}
here we used $\Delta$ as a low-frequency cut-off. 

We contrast Eq.~(\ref{eq:gamma3-bis}) with the one obtained for a weak anharmonic oscillator, after approximating $1-\cos\varphi\approx \varphi^2/2-\varphi^4/24$ in the Josephson term of Eq.~\eqref{eq:H0} and neglecting the phase slips. In that case, the gate sensitivity is absent, the right-hand side of the identity (\ref{eq:phase-slip}) is zero, rendering $\gamma_{\rm si}(\omega'|\omega)=0$. In harmonic approximation, Eq.~\eqref{eq:amplitude} vanishes identically. Treating 
{the anharmonic corrections to $\varepsilon_s$ and $W_{sr}$ appearing in Eq.~\eqref{eq:amplitude} perturbatively,} and assuming the incident photon to be close to resonance, $|\omega-\omega_0|\ll \omega_0$, yields
\begin{equation}
\label{eq:cross-quartic}
\gamma_{\rm{q}}(\omega_1,\omega_2,\omega_3|\omega)=
\frac{({256}/{3\pi^2})\Gamma^4\tilde E_C^2\omega^3_0\omega_1\omega_2\omega_3}{[(\omega_0-\omega)(\omega_0^2-\omega_1^2)(\omega_0^2-\omega_2^2)(\omega_0^2-\omega_3^2)]^2}.
\end{equation}
Assuming that the outgoing photon is also close to resonance, $|\omega'-\omega_0|\ll \omega_0$, we get
\begin{equation}
 \label{eq:gamma3-quartic}
\gamma_{\rm{q}}(\omega'|\omega)=\frac{32}{9\pi^2}\frac{\Gamma^4\tilde E_C^2(\omega-\omega')^3}{\omega_0^6(\omega-\omega_0)^2(\omega'-\omega_0)^2}.
\end{equation}

The comparison of Eqs.~\eqref{eq:gamma3-bis} and \eqref{eq:gamma3-quartic} shows that phase slips are much more effective in coupling the resonant modes to the low-frequency ones than the anharmonic corrections to the qubit levels. The low-frequency modes, being far away from the resonance, do not hybridize well with the qubit.  The phase slips are free from that drawback (at the expense of a potentially small {value of} $\lambda_1$). Thus phase slips dominate in the inelastic processes at $\omega'\to\omega$.

{\it Differential cross-section.--} We proceed further by accounting for higher-order {quasielastic} processes at finite $K$. The dichotomy between the high-frequency photon modes that are in resonance with the transmon and the low-frequency modes motivates a two-band approximation:
\begin{eqnarray}
\label{eq:H2}
H_\mathrm{eff}&= &
\omega_0|1\rangle\langle 1|
+\sum_{p>p_c}\omega_p|p\rangle\langle p|
+\sum_{p>p_c} \left[ t |p\rangle\langle 1|+{\rm H.c.}\right]
\nonumber\\
&&+\sum_{0<p<p_c} \omega_p a^\dagger_p a_p
+\lambda_1   |1\rangle\langle 1| \cos(2\pi \tilde n).
\end{eqnarray}
Here the first line represents the hybridization of the transmon with high-frequency photons; $|p\rangle=a^\dagger_p|0\rangle$, where $|0\rangle$ is the ground state, $|1\rangle$ is the state in which (only) the transmon is excited, and $t=\sqrt{\Gamma \Delta/\pi}$ is the hybridization matrix element. The second line in Eq.~\eqref{eq:H2} accounts for the coupling of low-frequency photons and the transmon excited state through phase slips~{\cite{footnote-a}}; the local charge operator $\tilde n$ differs from Eq.~\eqref{eq:charge-displacement} by the restriction of the sum to low-frequency modes, $0<p<p_c$. The separation between low- and high-frequency photons is set by frequency $\omega_c=vp_c$ such that $\Gamma\ll \omega_0-\omega_c\ll\omega_0$. 

The first line in the Hamiltonian \eqref{eq:H2} is equivalent to the Fano-Anderson model. Its eigenstate $|k\rangle$ with energy $\omega_k$ such that $\omega_k-\omega_0=-\Gamma\tan({\omega_k d}/{v})$ has an overlap with the transmon state
\begin{equation}
\beta^2_k\equiv|\langle 1|k\rangle|^2=\frac{{{\Gamma \Delta/\pi}}}{(\omega_k-\omega_0)^2+\Gamma^2}\,,
\end{equation}
assuming $\omega_k$ is close to the resonance, $|\omega_k-\omega_0|\ll \omega_0$ \cite{footnote1}. In new variables, Hamiltonian \eqref{eq:H2} then reads
\begin{eqnarray}
\label{eq:Hinter-2}
H_\mathrm{eff}&=&\sum_{k>p_c}\omega_k|k\rangle\langle k|
+\sum_{0<p<p_c} {\omega_p} a^\dagger_pa_p+H_1,
\nonumber\\
H_1&=&\lambda_1   \sum_{k,k'>p_c} \beta_k\beta_{k'}|k\rangle\langle k'|\cos(2\pi \tilde n).
\end{eqnarray}

It is the ``backaction" of the qubit on the dynamic charge $\tilde n$ that leads to the emission of ``soft" photon modes by the resonant ones~{\cite{footnote8}}. Using Eq.~\eqref{eq:Hinter-2}, we apply Fermi's Golden Rule to calculate the {(quasielastic) fluorescence spectrum} perturbatively in $\lambda_1$,
\begin{equation}
\label{eq:rate-ps}
\gamma_{\rm si}(\omega_{{k'}}|\omega_{{k}})=\frac{2\pi^2}{\Delta^2}\sum_{f}\left|\langle k',f|H_1|k,0\rangle\right|^2\delta(\omega_{{k}}-\omega_{{k'}}-E_f).
\end{equation}
Here $|k,f\rangle=|k\rangle\otimes|f\rangle$ and $|f\rangle$ is a multiphoton state with energy $E_f$ formed out of low-frequency photon modes. By standard manipulations, we express Eq.~\eqref{eq:rate-ps} in terms of a photon correlation function for an array disconnected from a transmon, 
\begin{equation}
\label{eq:rate-ps2}
\gamma_{\rm si}(\omega'|\omega)=\frac{\lambda_1^2\Gamma^2/\pi}{[(\omega-\omega_0)^2+\Gamma^2][(\omega'-\omega_0)^2+\Gamma^2]}{\cal C}(\omega-\omega')
\end{equation}
with
\begin{equation}
\label{eq:correlator}
\!{\cal C}(\Omega)=2{\rm Re}\int_0^{\infty}\!\!\!dte^{i(\Omega +i0^+)t}\langle \cos 2\pi \tilde n(t)\cos 2\pi \tilde n(0)\rangle.
\end{equation}
Here $\tilde n(t)= (1/ \pi) \sum_{0<p<p_c}f_p(a_pe^{-i\omega_p t}+a_p^\dagger e^{i\omega_p t})$.
We use extensively Baker-Hausdorff formula~\cite{Baker} to find
\begin{equation}
\label{eq:rate-ps3}
{\cal C}(\Omega)
=2e^{-4\sum_pf_p^2}
{\rm Re}\int_0^{\infty}\!\!\!dte^{i(\Omega +i0^+)t}\cosh\left( 4\sum_pf_p^2e^{-i\omega_p t}\right)
\end{equation}
at zero temperature. The Taylor expansion of the $\cosh$-factor in Eq.~\eqref{eq:rate-ps3} allows interpreting Eq.~\eqref{eq:rate-ps2} as a partial cross-section of a high-frequency photon {scattering} into another high-frequency photon, while an even number of low-frequency photons is produced. At $K\to 0$, Eq.~\eqref{eq:rate-ps2} reproduces Eq.~\eqref{eq:gamma3-bis} upon the renormalization~\cite{footnote4} of the phase slip amplitude, $\lambda_1\to \lambda_1 e^{-2\sum_pf_p^2}\approx \lambda_1(\Delta/\omega_0)^{2 K}$, and not too close to the resonance, $\Gamma\ll |\omega-\omega_0|,|\omega'-\omega_0|\ll\omega_0$. 

Being proportional to $\lambda_1^2(\Delta/\omega_0)^{4 K}$, the three-photon amplitude vanishes in the thermodynamic limit at finite $K$. Thus higher-order processes should be included. Instead of evaluating and summing them we note that at $\Delta\to 0$ Eq.~\eqref{eq:correlator} can be simplified,
\begin{equation}
\label{eq:correlator2}
{\cal C}(\Omega)\approx{\rm Re}\int_0^{\infty}\!\!\!dte^{i(\Omega +i0^+)t}\langle e^{{i} 2\pi \tilde n(t)}e^{-{i}  2\pi \tilde n(0)}\,\rangle\,.
\end{equation}
This correlator has been much studied~\cite{Weiss1985,Schon1990},
\begin{equation}
\label{eq:correl}
{\cal C}(\Omega)=
\frac \pi{\Gamma(4K)}\frac 1\Omega\left(\frac{\Omega}{\omega_0}\right)^{4K}e^{-\Omega/\omega_0}\,,\quad {\Omega>0}\,.
\end{equation}
Here $\Gamma(4K)$ is the Gamma function. {Equation \eqref{eq:rate-ps2} with ${\cal C}(\Omega)$ of Eq.~\eqref{eq:correl} is our main result~{\cite{footnote-b}}. It relates the fluorescence spectrum with the dynamical phase-boost susceptibility $\propto{\cal C}(\omega)$ at $\omega\ll \omega_0$~\cite{Guinea1985} in the same range of validity defined by Eq.~\eqref{eq:condition}.

Inserting Eq.~(\ref{eq:correl}) in Eq.~\eqref{eq:rate-ps2} we find that at resonant excitation, $|\omega-\omega_0|\lesssim\Gamma$, and in the frequency range $\omega_0\gg\omega_0-\omega'\gg\Gamma$ of the emitted photons, the fluorescence intensity is a power-law~\cite{footnote5} of $\omega_0-\omega'$,
\begin{equation}
{\gamma_{\rm si}(\omega'|\omega)=\frac {1}{\Gamma(4K)}\frac{\lambda_1^2}{\omega_0^3}\left(\frac{\omega_0-\omega^\prime}{\omega_0}\right)^{4K-3}}\,.
\label{gammasi}
\end{equation}
The perturbative-in-$\lambda_1$ result for the differential cross-section of {quasielastic} scattering works at any $K$, except its smallest values allowing for ${\lambda_1\gtrsim\Gamma}$. {The behavior of Eq.~\eqref{gammasi} parallels the one of the dynamical susceptibility $A(\omega)$ of Eq.~(10) in Ref.~\cite{Guinea1985b}. These two quantities are not normalizable at $K<1/2$, which is the signature of the charge {localized} phase. At the critical point ($K=1/2$), we find $\gamma(\omega'|\omega_0)\propto 1/(\omega_0-\omega')$ as the dynamical critical signature of the Schmid transition.

{\it Total inelastic cross-section.--} Finally, we show that at $K<1/2$ the {quasielastic} transitions may yield the main contribution to the {\it total} inelastic cross-section~\cite{footnote6},
 \begin{equation}
 \label{eq:total-cross}
  \gamma(\omega)=\int_0^\omega d\omega'\gamma(\omega'|\omega)\,.
 \end{equation}
Indeed, at $K<1/2$ the dominant contribution of the partial cross-section \eqref{eq:rate-ps2} to the integral comes from a vicinity of order $\Gamma$ near its upper bound. We may thus extend the lower bound in Eq.~\eqref{eq:total-cross} to $-\infty$ and evaluate the {quasielastic} component of the total cross-section as
\begin{equation}
\label{inel1}
\gamma_{\rm si}(\omega_0)
=\frac \pi{ 2\sin(2\pi K) \Gamma(4K)}\frac{\lambda_1^2}{\Gamma^2}
\left(\frac\Gamma{\omega_0}\right)^{4K}
\end{equation} 
for the incoming photons lying within the width of the resonance. Furthermore, the inelastic spectral linewidth is asymmetric, see Fig.~\ref{F:inel-zeroT}, with asymptotes
\begin{equation}
\gamma_{\rm si}(\omega)=\frac{\pi\lambda_1^2\Gamma/\omega_0^3}{\Gamma(4K)}\left(\frac{\omega_0}{\omega-\omega_0}\right)^{3-4K}
\label{inel2}
\end{equation}
at $\omega-\omega_0\gg\Gamma$ and
\begin{equation}
\gamma_{\rm si}(\omega)=\frac{\pi(1-4K)\lambda_1^2\Gamma^2/\omega_0^4}{\sin(4\pi K)\Gamma(4K)}\left(\frac{\omega_0}{\omega_0-\omega}\right)^{4-4K}
\label{inel3}
\end{equation}
at $\omega_0-\omega\gg\Gamma$. Importantly, the found $\gamma_{\rm si}(\omega)$ is independent of the artificially-introduced partition frequency $\omega_{c}$ thus justifying the use of  Eq.~\eqref{eq:H2}.

To assess the contribution to Eq.~(\ref{eq:total-cross}) of deeply-inelastic processes, we use the differential cross-section (\ref{eq:cross-quartic}) stemming from the transmon anharmonicity and favoring large energy transfer between the incoming and any of the outgoing photons. The corresponding contribution to the total cross-section is
\begin{equation}
\gamma_{\rm q}(\omega_0)=\alpha\frac{\Gamma^2\tilde E_C^2}{\omega_0^4}\quad \mathrm{with} \quad \alpha\approx 0.45.
\end{equation}
The comparison with Eq.~\eqref{inel1} shows that {quasielastic} processes dominate the total inelastic cross-section $\gamma(\omega)$ if $\lambda_1/\tilde E_C\gg (\Gamma/\omega_0)^{2(1-K)}$. Under the assumption $\lambda_1\ll\Gamma$, this condition is possible to satisfy only at $K<1/2$: in terms of the Schmid transition, the phase-slip mechanism may dominate the total inelastic cross-section only in the charge-localized phase.

While Eqs.~\eqref{eq:rate-ps2} and \eqref{eq:correl} remain valid at $K>1/2$, their use in evaluation of $\gamma(\omega)$ is not justified: the dominant contribution to the integral in Eq.~(\ref{eq:total-cross}) at $K>1/2$ comes from $\omega-\omega^\prime\gtrsim\omega_c$ and  depends on $\omega_c$, rendering the model (\ref{eq:H2}) inapplicable. This is consistent with our perturbative analysis of Eq.~(\ref{eq:H}): there is no parameter allowing to single out the phase-slip-induced transitions from other processes at energy losses comparable to $\omega_0$.

\begin{figure}
\includegraphics[width=0.9\columnwidth]{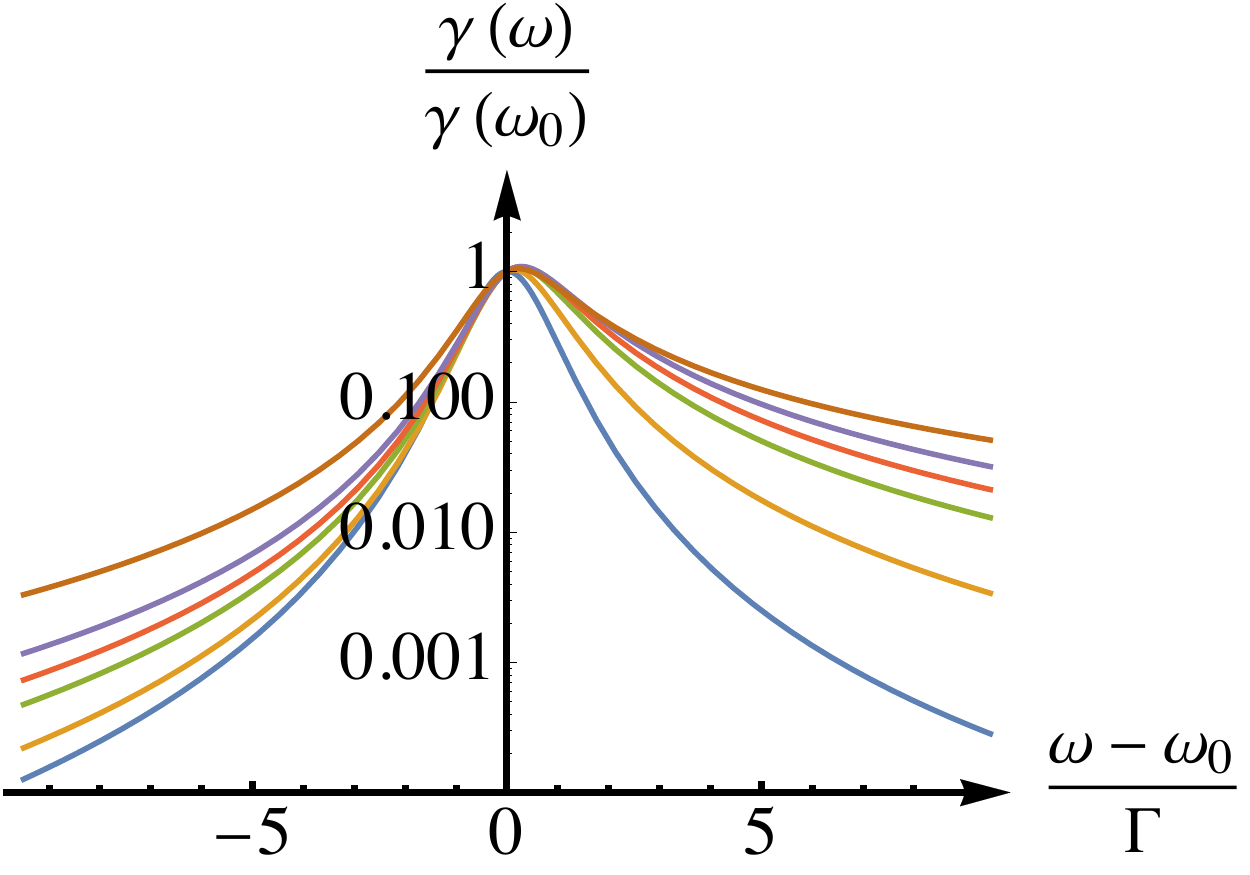}
\caption{
\label{F:inel-zeroT} 
Frequency dependence of the total cross-section at zero temperature for $K=0.01$ (blue), 0.1 (orange), 0.2 (green), 0.25 (red), 0.3 (purple), and 0.4 (brown).}
\end{figure}

It is straightforward to generalize Eqs.~(\ref{eq:rate-ps2}) and (\ref{inel1})-(\ref{inel3}) to finite temperatures $T$, by using the corresponding finite-$T$ generalization~\cite{Weiss1985,Schon1990} of ${\cal C}(\Omega)$. A low temperature, $T\ll\omega_0$, leaves the power-law spectrum~(\ref{gammasi}) intact at $\omega_0-\omega^\prime\gg T$. 
The total inelastic cross-section in the scaling region [$\Gamma,|\omega-\omega_0|,T\ll\omega_0$]  is found as
\begin{equation}
\label{eq:inel-final-T}
\frac{\gamma(\omega,T)}{\gamma(\omega_0,0)}=
\frac{\sin(2\pi K)}{\pi^2} \frac{\tau^{4K-1}}{1+\nu^2}\int_0^\infty dx
\frac{e^{\pi x/\tau}|\Gamma(2K+i x/\tau)|^2}{(x-\nu)^2+1}
\end{equation}
with $\tau=2\pi T/\Gamma$ and $\nu=(\omega-\omega_0)/\Gamma$. Its temperature dependence at  resonance, $\omega=\omega_0$, is shown in Fig.~\ref{F:inel}. 
The total cross-section increases (decreases) with the temperature at $K>1/4$ ($K<1/4$). 
At low temperature $T\ll \Gamma$,
\begin{equation}
\label{eq:inel-final-lowT}
\frac{\gamma(\omega_0,T)}{\gamma(\omega_0,0)}\approx1-\frac{1}{2}\left(\frac{2\pi T}{\Gamma}\right)^{4K},
\end{equation}
where the second term is a significant correction in a wide temperature range at $K\ll 1$.

\begin{figure}
\includegraphics[width=0.9\columnwidth]{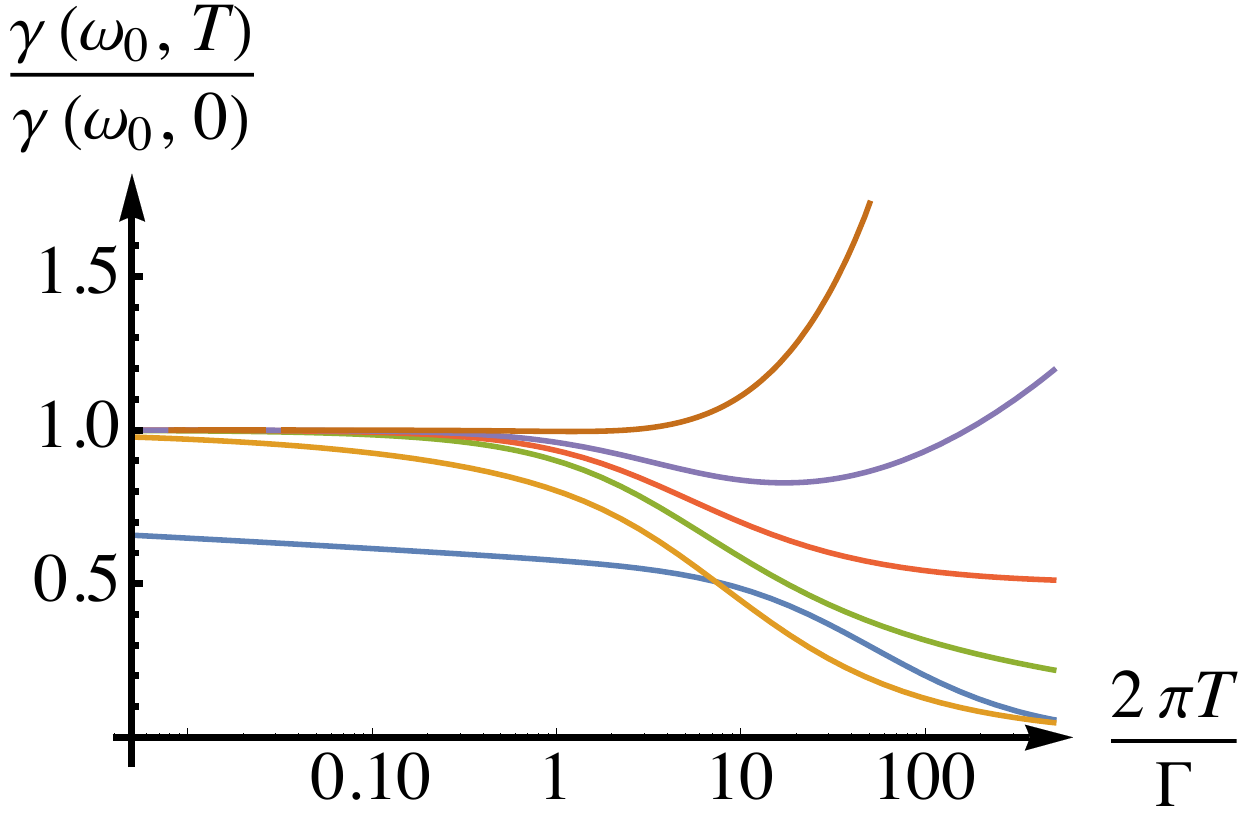}
\caption{
\label{F:inel} 
Temperature {dependence} of the total cross-section at resonance (same color code as in Fig.~\ref{F:inel-zeroT}).
}
\end{figure} 

{\it Conclusion.--} We believe the spectrum of the fluorescence {that we predict, see} Eqs.~(\ref{eq:rate-ps2})-(\ref{gammasi}), charts an interesting direction for future experiments, while the found total inelastic cross-section, see Eqs.~(\ref{eq:total-cross})-(\ref{eq:inel-final-lowT}), is directly related to the ongoing experiments~\cite{Kuzmin2020} in the spirit of Refs.~\cite{Puertas2019,Kuzmin2019,Leger2019}. Indeed, the internal quality factor $Q(\omega)$ of a discrete mode in a finite-length array, which is routinely measured in such experiments, can be expressed in terms of an inelastic decay rate, $Q(\omega)=\omega/\Gamma_{\rm in}(\omega)$. The latter is related to the total inelastic cross-section through $\Gamma_{\rm in}(\omega)=\gamma(\omega)\Delta/\pi$. Lastly, in the weakly nonlinear transmon regime, which we focussed upon, photon scattering remains mostly elastic. As the nonlinearity increases, we may anticipate large inelastic cross-sections, which would manifest a different kind of quantum impurity problem than the Kondo regime studied in Refs.~\cite{LeHur2012,Goldstein2013}.

\acknowledgments

We acknowledge stimulating discussions with M.~Goldstein and with V.~Manucharyan who attracted our attention to the role of phase slips{, as well as B. Beri and N. Roch for useful comments on the manuscript}. This work is supported by the DOE contract DE-FG02-08ER46482 (LG), and by ANR through Grant No.~ANR-16-CE30-0019 and ARO grant W911NF-18-1-0212 (MH).


\begin{thebibliography}{50}
\bibitem{footnote4}
In this limit, the intermediate scale $\omega_c$ can be replaced with $\omega_0$, and thus drops from the result.
\end{thebibliography}
\end{document}